\documentclass[runningheads]{llncs}
\usepackage[T1]{fontenc}
\usepackage{graphicx}
\usepackage{todonotes}
\usepackage{amsmath}
\usepackage{booktabs}

\begin{document}
\title{Physics-Informed Implicit Neural Representations for Improved Myocardial Perfusion MRI Quantification}
\titlerunning{Physics-Informed INRs Improve Myocardial Perfusion MRI Quantification}
%

\author{Christos Tsepas\inst{1} \and
Chang Yan\inst{2} \and
Maximilian Fuetterer\inst{2} \and
Sebastian Kozerke\inst{2} \and
Cian M. Scannell\inst{1}}

\institute{Department of Biomedical Engineering, Eindhoven University of Technology, Eindhoven, the Netherlands \and
Institute for Biomedical Engineering, University and ETH Zurich, Zurich, Switzerland\\
\email{c.m.scannell@tue.nl}}

\authorrunning{C. Tsepas et al.}
%
%
\maketitle              
\begin{abstract}
Quantifying myocardial perfusion from cardiac magnetic resonance (CMR) can be achieved by fitting tracer-kinetic models to the dynamic contrast-enhanced MR data. However, fitting the observed data with multi-compartment exchange models, which describe the evolution of the contrast agent in the tissue, to estimate perfusion parameters is a challenging inverse problem that is sensitive to noise
and acquisition variability. Previously, physics-informed neural networks (PINNs) have been proposed as an alternative to conventional non-linear least squares fitting methods with promising results for quantitative perfusion CMR. In this work, we extend the previously proposed PINN framework with spatiotemporal implicit neural representations (INRs) to represent the MR signal as a continuous spatiotemporal function and to improve the accuracy, smoothness, and physical consistency of the PINN model.
In realistic simulated CMR datasets, our proposed PINN with INRs demonstrates improved robustness and parameter estimation accuracy over the previously established methods. The code is available at \url{https://github.com/q-cardIA/pinn-inr}.

\end{abstract}
\section{Introduction}
\begin{figure}
    \centering
    \includegraphics[width=\linewidth]{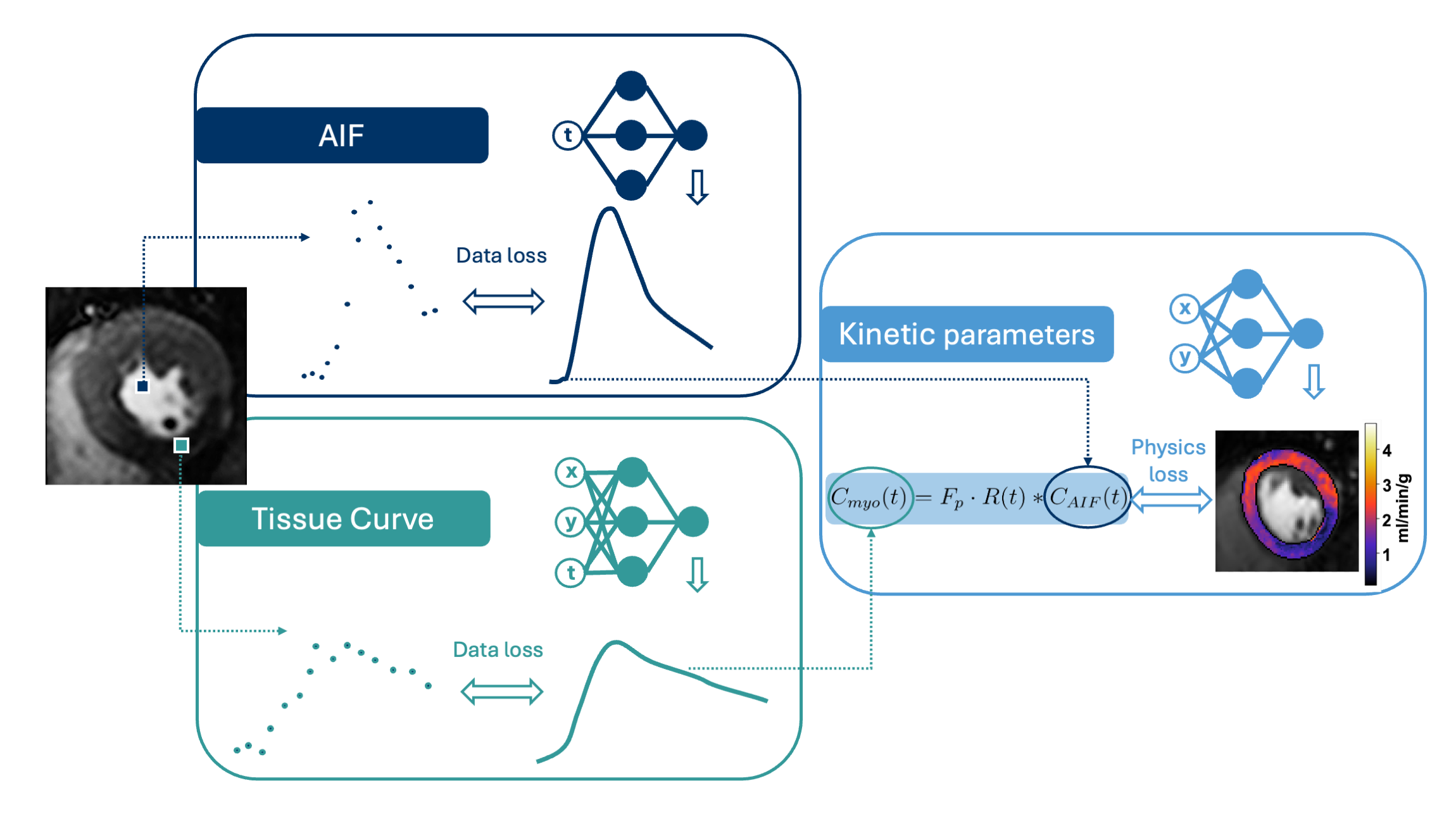}
    \vspace{-3mm}
    \caption{Physics-informed neural network modelling framework. The framework consists of three INRs: two (left) learn representations of the contrast concentration in the arterial input function (AIF) and myocardial tissue, respectively, guided by the data loss with the observed data. The third (right) learns the kinetic parameters of the tissue. The kinetic parameters are optimised to minimise the residual of the tracer-kinetic models (depending on the learned concentration values and their derivatives).}
    \vspace{-5mm}
    \label{fig:pinn-overview}
\end{figure}

Stress perfusion cardiac magnetic resonance (CMR) plays an important role in assessing myocardial ischaemia and is becoming established in clinical routine for the investigation of a wide-range of cardiovascular diseases. Quantitative perfusion CMR, in which absolute myocardial blood flow (MBF) values are derived from contrast agent signal dynamics, offers a more objective and reproducible alternative to visual assessment \cite{consensus}. It has been shown to improve diagnostic accuracy in the detection of multi-vessel and microvascular disease, where relative visual assessment is known to fail \cite{patel2010assessment}, \cite{rahman2019coronary}. 
Despite its potential, quantitative perfusion CMR still faces challenges related to the modelling pipeline, sensitivity to motion, artefacts, and noise.
Quantitative myocardial perfusion analysis relies on tracer-kinetic modelling to describe the dynamics of the gadolinium-based contrast agent within the tissue. These models provide a mathematical framework that links the measured MRI signal with underlying physiological parameters, such as blood flow and tissue volumes \cite{ingrisch2013tracer}. Among these models, the two-compartment exchange model (2CXM) is widely adopted for myocardial perfusion quantification and has the ability to capture the exchange between the plasma and interstitial spaces \cite{jerosch2010quantification}. Conventionally, model parameters are estimated voxelwise using non-linear least squares (NLLS) fitting. However, this approach is sensitive to noise, artefacts, and treats each voxel independently, thereby neglecting the spatial coherence of the underlying tissue properties.
Physics-informed neural networks (PINNs) have recently emerged as a promising alternative to NLLS fitting for quantitative MRI \cite{raissi2017physics}, \cite{banerjee2024pinns}. By embedding the governing tracer-kinetic equations directly into the loss function, PINNs enable self-supervised parameter estimation that is robust to low signal-to-noise ratio (SNR), limited temporal resolution, and short acquisition windows \cite{van2022physics}. In quantitative perfusion CMR, PINNs have demonstrated encouraging results in both simulations and patient data. However, existing formulations operate voxelwise: they model the temporal dynamics of the contrast agent passage but take no spatial input, so correlations between neighbouring voxels are not exploited and the estimated parameter maps are not regularised across space, which limits their smoothness and physiological consistency. \\
\indent Implicit neural representations (INRs) provide a complementary framework in which signals are modelled as continuous functions mapping spatial and temporal coordinates to signal values \cite{dupont2022data}. When combined with periodic activation functions, such as sinusoidal representation networks (SIRENs), INRs can represent high-frequency spatiotemporal signals with high fidelity and memory efficiency \cite{sitzmann2020implicit}. This continuous formulation is particularly well-suited to dynamic contrast-enhanced imaging, where the underlying signal is smooth in both space and time. INRs have been successfully applied to cerebral CT perfusion in acute ischaemic stroke, providing accurate estimates of perfusion parameters across a range of noise levels \cite{de2023spatio}. \\
\indent In this work, we integrate spatiotemporal implicit neural representations into the PINN framework for quantitative myocardial perfusion CMR, our central innovation being to represent the kinetic parameter maps as a continuous function of space, learned by a dedicated INR, rather than as independent per-voxel scalars as in previous PINN formulations \cite{van2022physics}, so that weight sharing across voxels acts as an implicit spatial regulariser. A separate INR represents the dynamic MR signal as a continuous function of space and time, providing a smooth and noise-robust signal estimate, and the INRs are coupled through the 2CXM, which is enforced as a physics constraint in the loss function (see Fig. \ref{fig:pinn-overview}). We further implement all networks as SIRENs, whose periodic activations overcome the spectral bias of conventional MLPs. We hypothesise that this combined formulation improves the accuracy, smoothness, and physical consistency of the estimated perfusion parameter maps relative to conventional NLLS fitting and previously proposed PINN approaches. Finally, we evaluate the method not only on a digital reference object but also on realistic CMR data generated by a Bloch-equation simulator, incorporating cardiac motion, providing a more challenging assessment than in previous work.

\section{Methods}
\subsection{Tracer-Kinetic Modelling}
As previously described, to quantify perfusion, a two-compartment exchange model is assumed to represent the physics of the transport process of a contrast agent across the tissue. That is, the perfusion system is assumed to be made up of two interacting compartments (plasma and interstitial fluid), and this gives rise to a pair of coupled ordinary differential equations (ODEs) which describe the evolution of the concentration of contrast agent over time in each compartment. The myocardium concentration $C_{MYO}$ is then given as the weighted sum of the concentrations in the separate compartments of $C_p$ and $C_e$,
\begin{equation}
    C_{MYO}(t) = v_p C_p(t) + v_e C_e(t),
\end{equation}
where $v_p$ and $v_e$ denote the fractional plasma and interstitial volumes, respectively, with $v_p + v_e \leq 1$. Solving the ODEs analytically shows that the tissue concentration can be written as a convolution of the arterial input function $C_{AIF}(t)$ with an impulse response function $R(t)$,
\begin{equation}
C_{MYO}(t) = F_p \cdot R(t) \ast C_{AIF}(t),
\end{equation}
where $F_p$ is the plasma flow and $R(t)$ depends on the four kinetic parameters $\eta = \{F_p, v_p, v_e,PS\}$, with $PS$ the permeability--surface area product. Quantitative perfusion analysis amounts to recovering $\eta$ voxelwise from the measured dynamic signal and a measured AIF, with the standard approach using non-linear least squares fitting algorithms to solve the inverse problem.

\subsection{PINNs with implicit neural representations}
\label{PINN-INR}
We build on the PINN framework of \cite{van2022physics}, in which a single multilayer perceptron (MLP), with tanh activations, takes the time coordinate $t$ as input and predicts $C_p, C_e$ and $C_{AIF}$ for all pixels in a slice, while the kinetic parameters $\eta = \{F_p, v_p, v_e, PS\}$ are treated as free per-pixel scalars that are optimised jointly with the network weights. This formulation has two limitations that motivate our extension. First, the network has no spatial input and predicts a separate column of outputs for every pixel, so spatial correlations between neighbouring voxels are not exploited and the kinetic parameter maps are not regularised across space. Second, the expressivity of a single MLP with tanh activations is limited. Such networks are biased towards low-frequency functions (the spectral bias of neural networks) and so tend to oversmooth the sharp contrast wash-in and inflection points that characterise perfusion concentration curves, which can in turn bias the recovered kinetic parameters.
 
As shown in Fig. \ref{fig:pinn-overview}, to address both limitations, we reformulate the problem using three coupled implicit neural representations (INRs). An INR models a signal as a continuous function from input coordinates (spatial and/or temporal) to signal values, with the network weights themselves encoding the signal \cite{sitzmann2020implicit}, \cite{dupont2022data}. In our case, the dynamic MR signal, the AIF, and the kinetic parameter maps are each represented as continuous functions of their natural coordinates. Crucially, the kinetic parameters are themselves modelled as a continuous spatial function rather than as free per-pixel scalars, which provides an implicit spatial regularisation of the recovered parameter maps. To overcome the spectral bias noted above, all three INRs are implemented as sinusoidal representation networks (SIRENs) \cite{sitzmann2020implicit}. The activation at hidden layer $\ell$ is
\begin{equation}
    \mathbf{a}^{(\ell)} = \sin\!\big(w_0 \cdot (\mathbf{W}^{(\ell)} \mathbf{a}^{(\ell-1)} + \mathbf{b}^{(\ell)})\big),
\label{eq:siren}
\end{equation}
where $\mathbf{W}^{(\ell)}$ and $\mathbf{b}^{(\ell)}$ are the learned weight matrix and bias of layer $\ell$, and $w_0$ is a fixed scalar that scales the pre-activation. A separate $w_0^{\text{init}}$ is applied at the first layer so that the input coordinates are mapped directly to an appropriate frequency range. The value of $w_0$ controls the maximum frequency content the network can represent: larger $w_0$ biases the INR towards higher-frequency functions. SIRENs have been shown to represent high-frequency spatiotemporal signals with substantially higher fidelity than conventional MLPs and have previously been used to model cerebral CT perfusion data \cite{de2023spatio}.

\paragraph{Network architectures.} The AIF network represents the arterial input as a continuous function of time, $\hat{C}_{AIF}(t) = f_{AIF}(t; \theta_{AIF})$,
and is implemented as a three-layer SIREN MLP with 64 hidden units per layer and frequencies $w_0 = w_0^{\text{init}} = 1$, reflecting the relatively smooth, low-frequency shape of the arterial input curve. The myocardium signal network represents the dynamic tissue concentration as a continuous function of space and time,
    $\hat{C}_{MYO}(\mathbf{x}, t) = f_{MYO}(\mathbf{x}, t; \theta_{MYO})$,
with four layers of 128 hidden units and SIREN frequencies $w_0 = w_0^{\text{init}} = 20$, chosen to capture the sharper wash-in and inflection points present in the tissue concentration curves and the spatial heterogeneity across the myocardium. The parameter INR maps spatial coordinates to the four kinetic parameters $[F_p(\mathbf{x}), v_e(\mathbf{x}), v_p(\mathbf{x}), PS(\mathbf{x})]^\top = f_{\eta}(\mathbf{x}; \theta_{\eta})$
and uses the same four-layer, 128-hidden-unit architecture as the myocardium network, but with lower SIREN frequencies $w_0 = 5$, $w_0^{\text{init}} = 10$, reflecting the lower spatial frequency content of the underlying parameter maps relative to the dynamic signal. To enforce the physical constraint that all kinetic parameters are strictly positive, the network predicts the logarithm of each parameter, and the exponential of the predictions are taken before being passed into the 2CXM forward model. The kinetic parameter INR is the central novel component of the proposed method: rather than treating the kinetic parameters as $4 \times |\Omega|$ independent scalars, they are constrained to lie on the continuous spatial manifold defined by $f_\eta$, which shares weights across all voxels and so naturally promotes spatially coherent parameter maps.
\paragraph{Loss function.} The framework is trained by minimising a weighted combination of data and physics losses. The data losses measure agreement between the INR predictions and the observed AIF and myocardium curves separately,
\begin{align}
    \mathcal{L}_{AIF} &= \frac{1}{|T|}\sum_{t \in T} \lVert \hat{C}_{AIF}(t) - C_{AIF}(t) \rVert^2, \\
    \mathcal{L}_{MYO} &= \frac{1}{|\Omega||T|} \sum_{\mathbf{x} \in \Omega} \sum_{t \in T} \lVert \hat{C}_{MYO}(\mathbf{x}, t) - C_{MYO}(\mathbf{x}, t) \rVert^2,
\end{align}
where $\Omega$ is the spatial domain and $T$ the set of acquisition times. The physics loss couples the three INRs through the 2CXM forward model in convolution form,
\begin{equation}
    \mathcal{L}_{\text{phys}} = \frac{1}{|\Omega||T|} \sum_{\mathbf{x} \in \Omega} \sum_{t \in T} \lVert \hat{C}_{MYO}(\mathbf{x}, t) - F_p(\mathbf{x}) \cdot R(t; \eta(\mathbf{x})) \ast \hat{C}_{AIF}(t) \rVert^2,
\end{equation}
where the parameters $\eta(\mathbf{x})$ entering the impulse response $R$ are produced by the parameter INR. In contrast to \cite{van2022physics}, who enforce the 2CXM as ODE residuals at randomly sampled collocation points using automatic differentiation, we enforce the equivalent analytical convolution form directly, which removes the need for collocation sampling and avoids the auxiliary $C_p$, $C_e$ outputs that are not separately observable in DCE-MRI.
 
The total loss is a weighted combination of $\mathcal{L}_{AIF}$, $\mathcal{L}_{MYO}$, and $\mathcal{L}_{phys}$
with weights $\lambda_{AIF} = 1.0$, $\lambda_{MYO} = 2.0$ and $\lambda_{\text{phys}} = 2.0$. All three networks are trained jointly with Adam (learning rate $10^{-4}$, batch size 16) for 900 epochs. To prevent the data terms from continuing to dominate the joint optimisation after they have converged, the AIF and myocardium networks are frozen at 75\% of the total training budget, after which only the parameter INR continues to be updated through the physics loss.

The SIREN frequency $w_0$ sets the frequency support of the represented function \cite{sitzmann2020implicit} and was therefore matched to the bandwidth of each signal: low for the smooth arterial input ($w_0 = 1$), high for the tissue curves with their sharper wash-in and spatial heterogeneity ($w_0 = 20$), and intermediate for the piecewise-smooth kinetic parameter maps ($w_0 = 5$). The loss weighting places the three terms on a comparable scale, so the AIF network, which fits a single rapidly converging curve, is given the lowest weight, while the myocardium data and physics terms are weighted equally. Because the data losses converge considerably faster than the physics loss, freezing the data networks at 75\% of the training budget acts as a two-stage optimisation in which the remaining budget is dedicated to the parameter INR under a fixed signal estimate. The code is made publicly available.\footnote[1]{\url{https://github.com/q-cardIA/pinn-inr}}

\subsection{Data}
\label{data}
The proposed method is developed and evaluated on two simulated datasets: a digital reference object (DRO) and realistic CMR simulations. Both provide ground-truth kinetic parameter maps, which are not available in patient data.

\subsubsection{Digital reference object}
\label{DRO}
The DRO follows the configuration of \cite{van2022physics}, \cite{Debus2019} to allow direct comparison. Known kinetic parameter values are used to forward simulate signal time curves via the 2CXM, producing a volume with ground-truth perfusion parameters at every location. The kinetic parameters are varied on a discrete grid of spatial dimensions $40 \times 120 \times 3$, where each $10\times10\times1$ block of voxels shares a unique combination of parameter values, yielding 144 distinct parameter combinations across the volume. The parameters and their discrete ranges are as follows: $Fp\in \{0.5,1.0,1.5,2.0\}$ ml/min/ml, $v_p \in \{0.02, 0.05, 0.1, 0.2\}$, $v_e\in \{0.1, 0.2, 0.5\}$, and $PS \in \{0.5, 1.5, 2.5\}$ ml/min/ml. Time curves are simulated over a window of 2 minutes at 100 time points, with the arterial input function (AIF) modelled using a gamma-variate function. Gaussian noise is added to the simulated curves at a signal-to-noise ratio of 17.5, consistent with levels observed in clinical myocardial perfusion acquisitions.

\subsubsection{Simulated CMR data}
\label{Simulated CMR}
Realistic DCE-CMR datasets are simulated using a particle-based Bloch equation simulator designed to accurately model motion and contrast changes in DCE-CMR \cite{Yan2026}. The perfusion simulation framework (CMRperf) takes an MRI sequence description, a numerical phantom, and a contrast model as input. The sequence was defined using the CMRseq framework \cite{cmrseq}. The extended cardiac-torso (XCAT) phantom provides anatomical masks with cardiac and/or respiratory vector motion fields \cite{xcat}, \cite{mrxcat}. Different contrast kinetic models can be used. Here, the 2CXM was deployed to simulate healthy and iscahemic myocardium \cite{jerosch2010quantification}, with the same gamma-variate AIF as used in the DRO.

\paragraph{Simulator design}
\label{Simulator design}
The simulator takes the XCAT numerical phantom, adding the pre-defined motion fields and contrast model, to generate a set of spin particles with time-varying positions and relaxation properties. In a first step, the XCAT motion vectors were used to generate a parametrisable deformation field with different motion types and scales. Approximately 27 randomly distributed spin particles with time-varying position and relaxation properties were simulated in each 1.5$\times$1.5$\times$1.5 mm$^3$ voxel space, with over 1 million particles in total. The moving spin particle field, along with the sequence waveform from CMRseq, was then fed into a Bloch simulator to generate k-space signals. To this end, a custom CUDA-based variant of CMRsim \cite{cmrsim} was extended to handle pre-defined motion fields and time-varying relaxation changes, such that the effect of motion and contrast change is reflected.

\paragraph{Simulated GRE sequence}
\label{Simulated GRE}
A typical 3-slice perfusion sequence with 120 ms saturation delay, 1.5$\times$1.5 mm$^2$ in-plane resolution, 8 mm slice thickness and spoiled GRE readout with 15-degree flip angle, 2.6 ms repetition time (TR), 1.39 ms echo time (TE) was used as an example. Rest perfusion of 1.5 ml/min/g at a heart rate of 60 beats/min for a total scan duration of 2 min was simulated. Prospective Cartesian random undersampling at R = 2,3, and 5 was performed and reconstructed using local-low rank (LLR) reconstruction \cite{llr}. MR signals were converted to gadolinium concentrations using sequence-based look-up tables. The typical simulation time is 4 hours on a NVIDIA RTX PRO 6000 Blackwell Workstation Edition.

\subsubsection{Evaluation}
\label{evaluation}
In both datasets, quantitative assessment is performed by comparing estimated parameter maps against the known ground-truth kinetic parameters using the normalised mean square error (NMSE) and the structural similarity index measure (SSIM). For the simulated CMR data, evaluation is performed across two different settings to compare performance: (i) with uniform kinetic parameters throughout the myocardium, and (ii) including a simulated lesion with reduced perfusion introduced in the anterolateral wall. The proposed method is compared versus the previous method \cite{van2022physics} and non-linear least squares (NLLS) fitting by fitting the 2CXM directly to the measured concentration time curves using the L-BFGS-B algorithm \cite{byrd1995limited} via \texttt{scipy.optimize.minimize}.

\section{Results}
\subsection{DRO}
Table~\ref{DRO_results} reports NMSE and SSIM for each of the four kinetic parameters across NLLS fitting, the voxelwise PINN of \cite{van2022physics}, and the proposed method (ST-PINN). Averaged across all parameters, ST-PINN achieved the lowest NMSE of 0.09 (0.14) and the highest SSIM of 0.63 (0.19), compared to 0.13 (0.09) and 0.53 (0.15) for PINN, and 0.17 (0.18) and 0.47 (0.14) for NLLS. ST-PINN performed particularly strongly on $F_p$ and $PS$, where it outperformed both baselines on NMSE by a considerable margin. Results were more mixed for $v_p$ and $v_e$, where the voxelwise PINN achieved lower NMSE, though ST-PINN consistently achieved higher SSIM across these parameters, suggesting more spatially coherent estimates.

\begin{table*}[htbp!]
    \centering
    \resizebox{\textwidth}{!}{
    \begin{tabular}{lcccccccc}
        \toprule
         & \multicolumn{2}{c}{$F_p$} & \multicolumn{2}{c}{$v_p$} & \multicolumn{2}{c}{$v_e$} & \multicolumn{2}{c}{$PS$} \\
        \cmidrule(lr){2-3} \cmidrule(lr){4-5} \cmidrule(lr){6-7} \cmidrule(lr){8-9} 
         & \text{NMSE} & \text{SSIM} & \text{NMSE} & \text{SSIM} & \text{NMSE} & \text{SSIM} & \text{NMSE} & \text{SSIM} \\
        \midrule
        \text{NLLS} & 0.03 (0.02) & 0.62 (0.03) & 0.07 (0.07) & 0.34 (0.15) & 0.02 (0.02) & 0.43 (0.25) & 0.55 (0.61) & \textbf{0.47 (0.14)} \\   
        \text{PINN} & 0.19 (0.09) & 0.75 (0.19) & \textbf{0.04 (0.02)} & 0.47 (0.11) & \textbf{0.01 (0.01)} & 0.58 (0.12) & 0.30 (0.25) & 0.34 (0.16) \\
        \text{ST-PINN} & \textbf{0.01 (0.002)} & \textbf{0.80 (0.07)} & 0.26 (0.2) & \textbf{0.58 (0.12)} & 0.03 (0.02) & \textbf{0.72 (0.09)} & \textbf{0.05 (0.03)} & 0.42 (0.16) \\

        \bottomrule
        \vspace{0.2em}

    \end{tabular}
    }
    \caption{Quantitative comparison of NLLS, PINN \cite{van2022physics}, and the proposed ST-PINN on the DRO. NMSE and SSIM are reported for each of the four kinetic parameters. Values are reported as mean (standard deviation). Bold indicates the best performing method for each metric. Lower NMSE and higher SSIM indicate better performance.}
    \label{DRO_results}
\end{table*}

\begin{figure}[htbp!]
    \centering
    \includegraphics[width=1\linewidth]{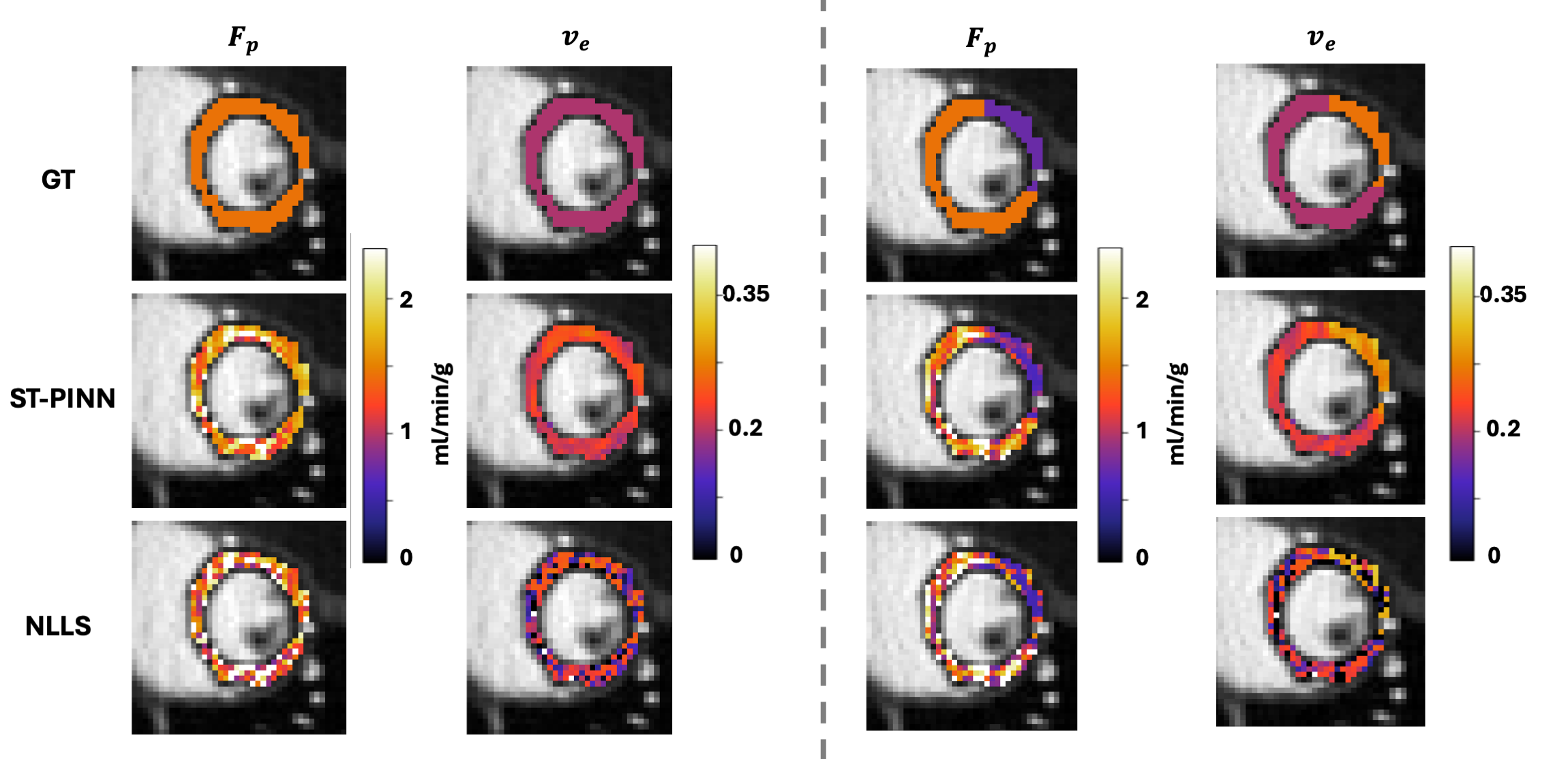}
    \caption{Qualitative comparison of $F_p$ and $v_e$ parameter maps estimated by ST-PINN and NLLS against the ground truth (GT) for a healthy (left) and ischaemic (right) simulated CMR case. The scattered values indicate the difficulty of the voxelwise fitting problem under noise and artefacts, and are more frequent in the NLLS maps than in the ST-PINN maps}
    \label{fig:results_fits}
\end{figure}

\subsection{CMRsim}
ST-PINN outperformed NLLS across all four kinetic parameters, with an overall median NMSE of 0.212 compared to 0.884 for NLLS. The improvement was most pronounced for $v_p$ and $v_e$, where NLLS exhibited substantially higher error and greater variability, suggesting that voxelwise fitting struggles with these parameters in the presence of motion and undersampling artefacts. ST-PINN also showed considerably tighter interquartile ranges across all parameters (0.36 vs 0.46), indicating more consistent estimation across slices and tissue types.

Figure~\ref{fig:results_fits} shows qualitative parameter maps for $F_p$ and $v_e$ in a healthy (left) and diseased (right) case, compared against the ground truth. ST-PINN produced spatially smooth maps that closely follow the ground truth in both conditions, correctly capturing the perfusion deficit in the anterolateral wall. NLLS maps exhibited considerable noise and spatial incoherence, with erratic voxelwise estimates that obscure the underlying perfusion pattern, particularly for $v_e$.

\section{Discussion and Conclusion}
We have presented a spatiotemporal PINN framework for quantitative myocardial perfusion CMR that integrates both spatial and temporal information into the parameter estimation pipeline.
By using implicit neural representations with periodic activation functions, the proposed method moves beyond the voxelwise formulation of prior PINN approaches and exploits the spatiotemporal continuity of the myocardial signal and the spatial smoothness of myocardial kinetic parameters to constrain the estimation problem more effectively. On the DRO, the proposed method achieved the best overall performance, with the lowest mean NMSE (0.09, versus 0.13 for the voxelwise PINN and 0.17 for NLLS) and the highest mean SSIM (0.63, versus 0.53 and 0.47). The improvement was not uniform across parameters: ST-PINN substantially reduced the error in $F_p$ and $PS$, whereas the voxelwise PINN achieved lower NMSE for $v_p$ and $v_e$. ST-PINN nonetheless gave the highest SSIM for three of the four parameters, consistent with the spatial regularisation imposed by the parameter INR yielding more spatially coherent maps. The relatively large error in $v_p$ reflects that the plasma volume is the least identifiable of the four parameters at the temporal resolution simulated here, as its contribution to the tissue curve is confined to the early first pass. The CMRsim experiments demonstrated improved recovery of all four kinetic parameters in terms of NMSE and SSIM over conventional NLLS fitting.

The simulated CMR evaluation, which incorporates physically accurate cardiac motion, contrast kinetics, and prospective undersampling via a Bloch equation simulator, provides a particularly challenging and realistic assessment of the method and shows that the proposed ST-PINN is more robust.
This suggests that the spatiotemporal representation provides a meaningful inductive bias that is well matched to the structure of perfusion CMR data, where kinetic parameters vary smoothly across the myocardium and signal evolution is continuous in time. 

This work has several limitations. First, the method has been developed and evaluated entirely in simulation; although the Bloch-based CMR simulations reproduce cardiac motion, contrast kinetics and undersampling, validation in patients across a range of cardiac pathologies remains an essential next step. Second, the SIREN frequencies must be specified separately for the AIF, the dynamic MR signal and the kinetic parameter maps, and performance depends on these choices. In practice, however, these values need not be tuned per subject: they are determined by the bandwidth of the underlying signals, which is set by the acquisition protocol (such as the temporal resolution, spatial resolution and voxel size) rather than by pathology, and can therefore be fixed once for a given protocol. A practical selection strategy is to tune them on a digital reference object matched to the protocol, where ground truth is available, or, since the training objective is fully self-supervised, to select them from the data and physics losses on held-out time frames without requiring ground truth.

In conclusion, we have demonstrated that spatiotemporal PINNs offer a more accurate alternative to existing approaches for quantitative myocardial perfusion CMR. By embedding the physics of tracer-kinetic modelling within a spatiotemporal INR, we impose physically consistent constraints on the parameter estimation problem, improving robustness to the noise and sparse sampling that are inherent to clinical DCE-MRI acquisitions. Unlike purely data-driven approaches, the physical constraints act as a form of regularisation, guiding the network towards physiologically plausible solutions even when the measured signal is degraded by noise, motion, or undersampling artefacts. These results provide a strong foundation for future work towards robust, automated perfusion quantification.

\section*{Acknowledgements}
This publication is part of the project QP-GPT: A foundation model for quantitative perfusion MRI with file number NGF.1609.243.041 of the research programme AiNed XS Europe which is (partly) financed by the Dutch Research Council (NWO) under the grant \url{https://doi.org/10.61686/PBGHH11782}.

%
%
%
\newpage
\bibliographystyle{splncs04}
%

\end{document}